\documentclass[fleqn,10pt]{wlscirep}
\usepackage{graphicx}
\usepackage{epsfig}
\usepackage{epstopdf}
\usepackage{amsmath}
\usepackage{multirow}
\usepackage{longtable}
\title{Universal linear-temperature resistivity: possible quantum diffusion transport in strongly correlated superconductors}

\author[1,$\star$]{Tao Hu}
\author[1]{Yinshang Liu}
\author[2,$\dag$]{Hong Xiao}
\author[1]{Gang Mu}
\author[3,4,5]{Yi-feng Yang}

\affil[1]{State Key Laboratory of Functional Materials for Informatics, Shanghai Institute of Microsystem and Information Technology, Chinese Academy of Sciences, 865 Changning Road, Shanghai 200050, China}
\affil[2]{Center for High Pressure Science and Technology Advanced Research, Beijing, 100094, China}
\affil[3]{Beijing National Laboratory for Condensed Matter Physics and Institute of Physics, Chinese Academy of Sciences, Beijing 100190, China}
\affil[4]{Collaborative Innovation Center of Quantum Matter, Beijing 100190, China}
\affil[5]{School of Physical Sciences, University of Chinese Academy of Sciences, Beijing 100190, China}

\affil[$\star$]{thu@mail.sim.ac.cn}
\affil[$\dag$]{hong.xiao@hpstar.ac.cn}

\begin{abstract}
The strongly correlated electron fluids in high temperature cuprate superconductors demonstrate an anomalous linear temperature ($T$) dependent resistivity behavior, which persists to a wide temperature range without exhibiting saturation. As cooling down, those electron fluids lose the resistivity and condense into the superfluid. However, the origin of the linear-$T$ resistivity behavior and its relationship to the strongly correlated superconductivity remain a mystery. Here we report a universal  relation $d\rho/dT=(\mu_0k_B/\hbar)\lambda^2_L$, which  bridges the slope of the linear-$T$-dependent resistivity ($d\rho/dT$) to the London penetration depth $\lambda_L$ at zero temperature among  cuprate superconductor Bi$_2$Sr$_2$CaCu$_2$O$_{8+\delta}$  and  heavy fermion  superconductors CeCoIn$_5$, where $\mu_0$  is vacuum permeability, $k_B$ is the Boltzmann constant and $\hbar$ is the reduced Planck constant. We extend this scaling relation to different systems and found that it holds for other cuprate, pnictide and heavy fermion superconductors as well, regardless of the significant differences in the strength of electronic correlations, transport directions, and doping levels. Our analysis suggests that the scaling relation in strongly correlated superconductors could be described as a hydrodynamic diffusive transport, with the diffusion coefficient ($D$) approaching the quantum limit $D\sim\hbar/m^*$, where $m^*$ is the quasi-particle effective mass.
\end{abstract}
\begin{document}

\flushbottom
\maketitle

\thispagestyle{empty}

\section*{Introduction}

In quantum mechanics, the uncertainty principle gives rise to quantum fluctuations of the system that may impose some universal bound on its physical properties. Calculations based on the AdS/CFT (Anti de-Sitter/Conformal Field Theory ) have suggested a lower bound for the liquid viscosity, $\eta/s\geq\hbar/4\pi k_B$ \cite{adscft}, where $\eta$ is the shear viscosity and $s$ the entropy. Recent experiments also revealed a quantum bound $D_s\geq\hbar/m$ for the spin diffusivity $D_s$ in a strongly interacting Fermi gas \cite{Sommer,bardon}. Here $\hbar$ is the reduced Planck constant and $m$ is the mass of particles. It is therefore interesting to ask if such a lower bound may be realized in the electron transport of strongly correlated quantum critical systems.

One of the distinguished features of strongly correlated cuprate superconductors is the linear-temperature ($T$) dependent resistivity \cite{gurvitch}, which could extend to very high temperature \cite{gurvitch} and violate the Mott-Ioffe-Regel (MIR) limit \cite{Ioffe}. The linear relationship has also been observed in some heavy fermion superconductors \cite{Petrovic,Paglione}, starting at around the superconducting transition temperature $T_c$ and extending to high temperatures of about $10-20$ times of $T_c$. Many different  mechanisms have been proposed to explain the microscopic origin of the linear-$T$ resistivity behavior including quantum critical theories and the more exotic AdS/CFT calculations. On the other hand, recent experiment suggested that the linear-$T$ resistivity in different materials may share a similar scattering rate \cite{bruin}.

In the present work, we investigated the linear-$T$ resistivity in a number of strongly correlated superconductors and demonstrate a connection between its coefficient and the superfluid density responsible for the charge carrying in the superconducting state. We show that this can be understood by a diffusion transport of heavy quasi-particles whose diffusion coefficient approaches the quantum limit $D=\hbar/m^*$, where $m^*$ is the effective mass of the quasi-particles.

\section*{Results}

We start with the heavy fermion superconductor CeCoIn$_5$. Among all strongly correlated superconductors, CeCoIn$_5$ is remarkably  similar to the high $T_c$ cuprate superconductors in several aspects \cite{Thompson}. For example, it has also a two-dimensional Fermi surface \cite{FM1,FM2}, its superconducting phase is near to an antiferromagnetic phase \cite{knebel2004high,phase2,phase3}, and its superconducting gap has $d$-wave symmetry \cite{dwave1,xiao2008pairing,zhou2013visualizing}. Besides, CeCoIn$_5$ is one of the purest strongly correlated superconductors \cite{Petrovic,Paglione}, with a tunable linear-$T$ resistivity under modest applied pressure \cite{Sidorov, Bianchi, Seo, Park}. To examine its transport properties, we have therefore grown high quality CeCoIn$_5$ single crystal samples by an indium self-flux method \cite{Petrovic} and performed detailed transport measurements under pressure to avoid disorder related effects.

Figure 1(a) demonstrates the $T$-dependent resistivity curve of CeCoIn$_5$ under pressure from 0 GPa to 1.0 GPa. All of them exhibit a perfect linear-in-$T$ resistivity from around $T_c$ to about 20 K as indicated by the dashed lines. The inset of Fig. 1(a) shows the $T$-dependent resistivity of CeCoIn$_5$ up to 300 K. For comparison, we also plot in Fig. 1(b) the resistivity of Bi$_2$Sr$_2$CaCu$_2$O$_{8+x}$ from underdoped to overdoped regime with the oxygen contents from $x=$ 0.2135 to 0.27 \cite{Watanabe}. Figure 1(c) demonstrates the $d\rho /dT$ versus $\lambda^{2}_L$  for both compounds, using the experimental results for the penetration depth measured previously by muon spin spectroscopy \cite{Howald} and ac susceptibility \cite{Anukool}. We see remarkably that all the investigated samples fall on the same straight line described by $d\rho /dT=(\mu_0k_B/\hbar)\lambda^{2}_L$, with a coefficient that is determined entirely by the fundamental constants ($\mu_0$: the vacuum permeability; $k_B$: the Boltzmann constant; $\hbar$: the reduced Planck constant). This indicates a universal origin for the charge transport in both compounds.

The above relation between $d\rho /dT$ and $\lambda^{2}_L$ can be extended to various other strongly correlated superconductors with linear-$T$ resistivity. The data are summarized in Fig. 2 on a log-log scale. Most resistivity data were taken from experimental results on high-quality single crystal samples in order to obtain the intrinsic linear-in-$T$ coefficient. The values of the penetration depth were obtained by muon spin spectroscopy \cite{Uemura}, optical conductivity measurement  \cite{Homes1,Homes2} and some other techniques. Note that for superconducting thin films, the experimental magnetic penetration depth generally deviates from the London penetration depth $\lambda_L$  due to structural disorders in the films \cite{Tinkham,Gubin}. Even in high quality ultrathin films, there is a large difference in superfluid density between the film and the bulk materials with same $T_c$ \cite{Lemberger,Hetel}. Consequently all the data of the London penetration depth shown in Fig. 2 were taken only from bulk materials. It is worth noting that Fig. 2  also includes the transport data for cuprate superconductors along different transport directions, e.g., YBa$_2$Cu$_3$O$_{6.93}$ along the $a$, $b$ and $c$-axis. Cuprates generally exhibit a metallic in-plane resistivity but an insulating-like resistivity along the $c$-axis below certain temperature, which reflects the two-dimensional nature of the system. Correspondingly, the penetration depth along the $c$-axis is determined by a Josephson-coupling between superconducting layers \cite{Basov2,Dordevic,Shibauchi}, which is different from the in-plane one \cite{Homes1}. Thus it is amazing to observe that the same scaling relation holds true for both directions. Combining the data for all the strongly correlated superconductors summarized here, we see that the scaling, $d\rho/dT=(\mu_0k_B/\hbar)\lambda^{2}_L$, spans over several orders of magnitude. Note that the in-plane LSCO data in the extremely underdoped regime $0.07\leq p\leq 0.12$ demonstrates a systematic deviation from the scaling relationship as shown in Fig. 2. The deviation could be understood in terms of the complex competing phase, like charge density wave and pseudogap, which become significant in the underdoped regime.

The above scaling relation is consistent with several well-known experimental facts. First, considering the special case at $T=T_c$ and neglecting the residual resistivity, the scaling relation $d\rho /dT\propto\lambda^{2}_L$ gives the well-known Homes's law, $\sigma_cT_c\propto\lambda^{-2}_L$, where $\sigma_c$ is the $dc$ conductivity at $T_c$ \cite{Homes1}. Second, the Drude formula \cite{drude} is often used to describe the resistivity of conventional metals, $\rho=m^*/n_ne^2\tau$, where $m^*$ is the effective mass of the quasi-particles, $n_n$ is the carrier density of quasi-particles, $e$ is the charge of electrons, and $\tau$ is the relaxation time. If we naively match the Drude formula with the above scaling relation for a non-quasiparticle system and assume that the normal fluid and the superfluid are composed of the same charge carriers, $\lambda_L=(m^*/\mu_0n_ne^2)^{1/2}$, we obtain immediately a material-independent scattering rate $\tau^{-1}=k_BT/\hbar$ for all these strongly correlated superconductors. This is consistent with the universal scattering rate recently observed in the linear-in-$T$ resistivity region among good and "bad" metals \cite{bruin}. 
However, one can not take it for granted that the normal fluid in Drude model and superfluid in London equation are always the same. Actually, experiments showed that only part of normal carriers condensate into superfluid \cite{tanner1998superfluid}. In addition, the measurements of the London moment already revealed the mass of Cooper pairs are undressed and have twice of the electron's bare mass, regardless the conventional metal superconductors \cite{hildebrandt1964magnetic}, heavy fermion superconductors \cite{sanzari1996london} or cuprates \cite{verheijen1990measurement}, which is different from the effective mass in the Drude formula. These results suggest that the mass and carrier density of the superfluid ($n_s$) and the normal fluid ($n_n$) are different in strongly correlated superconductors. So one can not directly obtain the universal scaling relation simply by replacing $\rho$  with Drude model and $\lambda_L$ with London equation. The universal scaling relation $d\rho /dT=(\mu_0k_B/\hbar)\lambda^{2}_L$ has much deeper physics, which directly links the superfluid at zero temperature to the normal fluid responsible for the linear-in-$T$ resistivity in strongly correlated superconductors. It reveals an underlying relation between the superfluid and normal carriers: $n_s/m_e=n_n/m^*$. And indeed experimental evidence shows that about one of fourth normal carriers \cite{tanner1998superfluid} condensates into superfluid in optimal doped cuprates while the effective mass of optimum cuprates is about 3-4 times of the electron free mass \cite{Padilla,tanner1998superfluid}, which validate $n_s/m_e=n_n/m^*$.

The above result provides important information on the nature of the electron transport in the quantum critical regime. Recently, several experiments have shown that electrons in solid can exhibit hydrodynamic flows similar to a classical viscous liquid, if the electron fluid equilibrates by the electron-electron collisions \cite{Bandurin,Crossno,Moll}. Thus the electron transport in strongly correlated superconductors, where electron-electron interactions play a major role in the scattering processes, might in principle have a hydrodynamic description. Consequently, its linear-in-$T$ resistivity could be described by the well-known Einstein's relation \cite{Einstein}, an important law for the hydrodynamic transport, which states that the mobility ($\mu$) of a particle in a fluid is related to its diffusion coefficient ($D$), namely, $D=\mu k_BT$. Hence we have $\rho=k_BT/n_ne^2D$ and in the linear-in-$T$ regime, the diffusion coefficient $D$ must be a temperature-independent constant. Combining this and the scaling relation immediately yields $D=\hbar/m^*$, which is the quantum limit of the charge diffusion coefficient for the quasi-particles with an effective mass, $m^*$. This is one of the most important consequence of our observations. Actually, the quantum limit of the diffusion coefficient was recently observed in cold fermionic atomic gases in the unitary limit of scattering \cite{Sommer,bardon}. It implies that quantum diffusion transport might be a universal property of strongly correlated fermionic systems where the electron scatterings are so strong that the transport becomes highly incoherent. In fact, it was proposed recently that the transport in an incoherent metal is controlled by the collective diffusion of energy and charge \cite{diffusion1}, supporting the proposed scenario of quantum diffusion transport in the present work. Thus, the obtained scaling relation suggests the superfluid could also be governed by the quantum diffusion, since it connects the ground state with the normal state in the strongly correlated superconductors.

Our results also provide some insights on the nature of strongly correlated superconductivity, which is often born out of strongly correlated normal fluid in the quantum critical regime. Since the latter already approaches the quantum diffusion limit before it transits into the superfluid state, it implies a zero-point motion of the superfluid. Some people considered the quantum diffusion as a necessary condition for the presence of superfluid \cite{Mendelssohn,HIrsch}. In fact, the quantum diffusion might explain the Uemura results for superconducting transition temperatures. Y. Uemura {\sl et al.} observed that the underdoped cuprate superconductors exhibit a Bose-Einstein-condensation (BEC)-like superconducting transition but with a reduced transition temperature \cite{Uemura,uemura2}. Actually, the BEC generally occurs when the thermal de Broglie wavelength  $\lambda_{dB}$ is comparable to the distance between bosons, where $\lambda_{dB}$  characterizes a length scale within which the bosons can be regarded as quantum mechanical wave-packets. However, the quantum diffusion gives a new length scale $\xi_{Th}=\sqrt{D\tau}$ with $\tau=\hbar/k_BT$, which characterizes the length scale that carriers can travel before losing their quantum coherence. Since the diffusion length $\xi_{Th}=\sqrt{\hbar^2/m^*k_BT}$  is less than $\lambda_{dB}=\sqrt{2\pi\hbar^2/2m^*k_BT}$  of electron pairs under certain temperature, it makes $\xi_{Th}$ a new dephasing length to determine the $BEC$ temperature. Thus the BEC temperature ($T_B$) is reduced to $T_c/T_B=(\xi_{Th}/\lambda_{dB})^2=1/\pi$  as observed in the Uemura plot \cite{Uemura,uemura2}.

\begin{figure}
\centering
\includegraphics[trim=0cm 0cm 0cm 0cm, clip=true, width=1\textwidth]{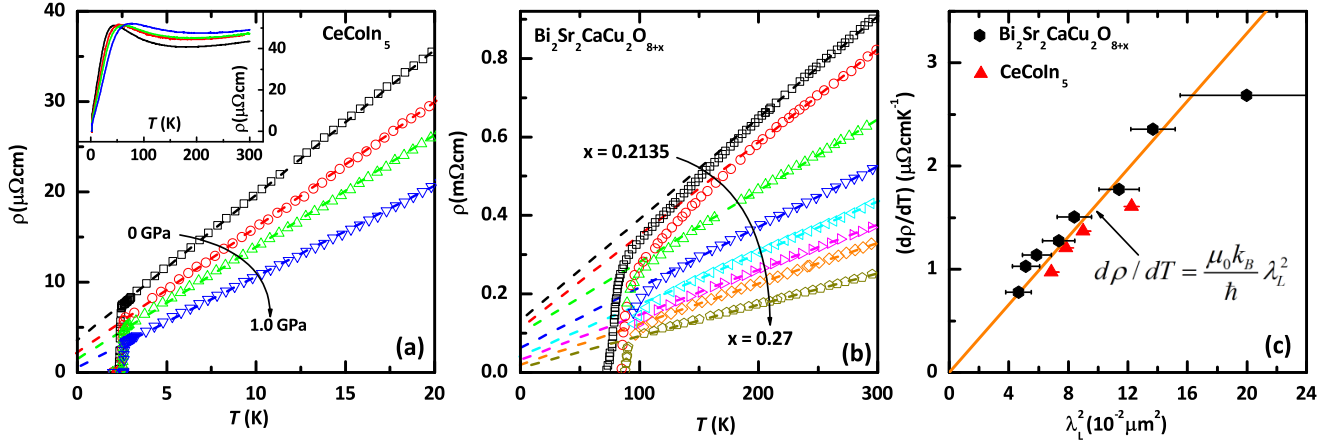}
\caption{\label{fig:TORQUEvsT} (Color online) (a) $T$ dependent resistivity $\rho$ of CeCoIn$_5$ under the pressure 0, 0.3, 0.55, 1.0 GPa. The arrow points to the increase in pressures. Inset to 1(a) is the  $\rho$ of CeCoIn$_5$ up to 300 K. (b) $T$ dependent resistivity $\rho$ of oxygen doped Bi$_2$Sr$_2$CaCu$_2$O$_{8+x}$ with $x=$ 0.2135, 0.217, 0.22, 0.24, 0.245, 0.255, 0.26, 0.27, respectively. The arrow points to the Bi2212 from underdoped to overdoped. The resistivity data of Bi$_2$Sr$_2$CaCu$_2$O$_{8+x}$ are taken from the literature \cite{Watanabe}. (c) Linear scale plot of $d\rho/dT$  vs. $\lambda^2_L$ for CeCoIn$_5$ (red triangles) and Bi$_2$Sr$_2$CaCu$_2$O$_{8+x}$ (Bi2212) (black circles). $d\rho/dT$  is the slope of linear-temperature-dependent resistivity, and  $\lambda_L$ is the London penetration depth of superconductors at zero temperature. The straight line corresponds to $d\rho/dT=(\mu_0k_B/\hbar)\lambda^2_L$, where $\rho$ is in the unit of $\mu\Omega cm$,  $\lambda_L$ is in $\mu m$, and $T$ is in K. See  Table 1 for details, including errors.  }
\end{figure}



\begin{small}
\begin{longtable}{|l|l|l|l|l|l|l|l|}
\hline
Material                    & label                                                                             & $T_c$(K)              & $d\rho/dT (\mu\Omega cmK^{-1})$ & \begin{tabular}[c]{@{}l@{}}$T$-linear range \\ in calculation (K)\end{tabular}                   & Ref.                                        & $\lambda_L(nm)$   & Ref.                      \\ \hline
$c$-axisYBa$_2$Cu$_3$O$_{6.93}$          & 1                                                                                 & 92                  & 12.3$\pm$0.07         & 250-400                                                                                        & \multirow{5}{*}{\cite{Takenaka}}                        & 923        & \multirow{3}{*}{\cite{Homes1}}       \\ \cline{1-5} \cline{7-7}
$c$-axisYBa$_2$Cu$_3$O$_{6.88}$          & 2                                                                                 & 92                  & 12.8$\pm$0.17         & 300-400                                                                                        &                                             & 1400       &                           \\ \cline{1-5} \cline{7-7}
$c$-axisYBa$_2$Cu$_3$O$_{6.78}$          & 3                                                                                 & 82                  & 30$\pm$0.84           & 320-400                                                                                        &                                             & 2900       &                           \\ \cline{1-5} \cline{7-8}
$a$-axis YB$_2$Cu$_3$O$_{6.93}$          & \multirow{4}{*}{YBa$_2$Cu$_3$O$_x$}                                                         & \multirow{4}{*}{92} & 0.78$\pm$0.001        & 105-300                                                                                        &                                             & 160        & \multirow{4}{*}{\cite{Tallon,Basov1}} \\ \cline{1-1} \cline{4-5} \cline{7-7}
$b$-axisYBa$_2$Cu$_3$O$_{6.93}$          &                                                                                   &                     & 0.37$\pm$0.004        & 150-300                                                                                        &                                             & 100        &                           \\ \cline{1-1} \cline{4-7}
$a$-axis YB$_2$Cu$_3$O$_7$             &                                                                                   &                     & 0.95$\pm$0.002        & 110-300                                                                                        & \multirow{2}{*}{\cite{Ando2}}                        & 160        &                           \\ \cline{1-1} \cline{4-5} \cline{7-7}
$b$-axisYBa$_2$Cu$_3$O$_7$             &                                                                                   &                     & 0.43$\pm$0.003        & 255-300                                                                                        &                                             & 100        &                           \\ \hline
Tl$_2$Ba$_2$CuO$_{6+x}$                & \multirow{2}{*}{Tl$_2$Ba$_2$CuO$_{6+x}$}                                                     & 80                  & 1.55$\pm$0.001       & 120-300                                                                                        & \cite{Tyler}                                         & 240$\pm$20    & \cite{Puchkov}                      \\ \cline{1-1} \cline{3-8}
$c$-axis Tl$_2$Ba$_2$CuO$_{6+x}$         &                                                                                   & 85                  & 2681$\pm$3.4          & 220-300                                                                                        & \cite{Schmann}                                         & 17500$\pm$2500 & \cite{Schmann,Tsvetkov,Moler}                  \\ \hline
Bi$_2$Sr$_2$CaCu$_2$O$_{8+x}$  ($x=$0.2135) & \multirow{8}{*}{Bi$_2$Sr$_2$CaCu$_2$O$_{8+x}$}                                                  & 72$\pm$2                & 2.68$\pm$0.01         & 200-300                                                                                        & \multirow{8}{*}{\cite{Watanabe,Matsuda}}                   & 447$\pm$50     & \multirow{8}{*}{\cite{Anukool}}      \\ \cline{1-1} \cline{3-5} \cline{7-7}
Bi$_2$Sr$_2$CaCu$_2$O$_{8+x}$  ($x=$0.217)  &                                                                                   & 79.3                & 2.36$\pm$0.02         & 200-300                                                                                        &                                             & 370$\pm$20    &                           \\ \cline{1-1} \cline{3-5} \cline{7-7}
Bi$_2$Sr$_2$CaCu$_2$O$_{8+x}$  ($x=$0.22)   &                                                                                   & 81.9                & 1.78$\pm$0.003        & 180-300                                                                                        &                                             & 338$\pm$20    &                           \\ \cline{1-1} \cline{3-5} \cline{7-7}
Bi$_2$Sr$_2$CaCu$_2$O$_{8+x}$  ($x=$0.24)   &                                                                                   & 87.9                & 1.5$\pm$0.002         & 160-300                                                                                        &                                             & 290$\pm$20    &                           \\ \cline{1-1} \cline{3-5} \cline{7-7}
Bi$_2$Sr$_2$CaCu$_2$O$_{8+x}$  ($x=$0.245)  &                                                                                   & 89                  & 1.28$\pm$0.002        & 155-300                                                                                        &                                             & 271$\pm$20    &                           \\ \cline{1-1} \cline{3-5} \cline{7-7}
Bi$_2$Sr$_2$CaCu$_2$O$_{8+x}$  ($x=$0.255)  &                                                                                   & 87.8                & 1.14$\pm$0.001        & 150-300                                                                                        &                                             & 243$\pm$20    &                           \\ \cline{1-1} \cline{3-5} \cline{7-7}
Bi$_2$Sr$_2$CaCu$_2$O$_{8+x}$  ($x=$0.26)   &                                                                                   & 86                  & 1.03$\pm$0.001        & 125-300                                                                                        &                                             & 227$\pm$20    &                           \\ \cline{1-1} \cline{3-5} \cline{7-7}
Bi$_2$Sr$_2$CaCu$_2$O$_{8+x}$  ($x=$0.27)   &                                                                                   & 84                  & 0.78$\pm$0.01         & 105-300                                                                                        &                                             & 216$\pm$20    &                           \\ \hline
SrFe$_2$(As$_{0.7}$P$_{0.3})_2$           & SrFe$_2$(As$_{0.7}$P$_{0.3})_2$                                                                 & 25                  & 0.7$\pm$0.01       & 100-300                                                                                        & \cite{Takahashi}                                         & 270$\pm$10    & \cite{Takahashi}                        \\ \hline
BaFe$_2$(As$_{0.67}$P$_{0.33})_2$         & BaFe$_2$(As$_{0.67}$P$_{0.33})_2$                                                               & 29.5                & 1.16$\pm$0.001        & 31-150                                                                                         & \cite{Kasahara}                                         & 315$\pm$15     & \cite{Hashimoto}                     \\ \hline
NaFe$_{0.97}$Co$_{0.03}$As            & NaFe$_{0.97}$Co$_{0.03}$As                                                                  & 21.8                & 1.46$\pm$0.004        & 50-250                                                                                         & \cite{Okada}                                         & 375$\pm$15     & \cite{Okada}                       \\ \hline
FeSe                        & FeSe                                                                              & 8                   & 5.84$\pm$0.01        & 20-80                                                                                          & \cite{Abdela}                                        & 425$\pm$20     & \cite{Hsu}                      \\ \hline
$c$-axis La$_{1.9}$Sr$_{0.1}$CuO$_4$       & 4                                                                                 & 27                  & 160$\pm$0.5           & 500-800                                                                                        & \multirow{4}{*}{\cite{Nakamura}}                        & 5908$\pm$400   & \multirow{30}{*}{\cite{Panagopoulos}}     \\ \cline{1-5} \cline{7-7}
$c$-axis La$_{1.9}$Sr$_{0.12}$CuO$_4$      & 3                                                                                 & 30                  & 195$\pm$0.2           & 300-800                                                                                        &                                             & 5345$\pm$400   &                           \\ \cline{1-5} \cline{7-7}
$c$-axis La$_{1.88}$Sr$_{0.15}$CuO$_4$     & 2                                                                                 & 35.8                & 154.8$\pm$0.3         & 300-800                                                                                        &                                             & 3816$\pm$280   &                           \\ \cline{1-5} \cline{7-7}
$c$-axis La$_{1.8}$Sr$_{0.2}$CuO$_4$       & 1                                                                                 & 31.7                & 147.6$\pm$0.2         & 320-800                                                                                        &                                             & 2441$\pm$200   &                           \\ \cline{1-7}
La$_{1.85}$Sr$_{0.15}$CuO$_4$            & \multirow{17}{*}{\begin{tabular}[c]{@{}l@{}}La$_{2-x}$Sr$_x$CuO$_4$\\ - Hussey\end{tabular}} &                     & 1.7$\pm$0.4           & \multicolumn{2}{l|}{\multirow{10}{*}{\begin{tabular}[c]{@{}l@{}}The slope are directly \\ taken from \\ Hussey et. al.\cite{Hussey}, \end{tabular}}} & 249$\pm$20     &                           \\ \cline{1-1} \cline{3-4} \cline{7-7}
La$_{1.84}$Sr$_{0.16}$CuO$_4$            &                                                                                   &                     & 1.5$\pm$0.4           & \multicolumn{2}{l|}{}                                                                                                                        & 229$\pm$20     &                           \\ \cline{1-1} \cline{3-4} \cline{7-7}
La$_{1.83}$Sr$_{0.17}$CuO$_4$            &                                                                                   &                     & 1.4$\pm$0.4           & \multicolumn{2}{l|}{}                                                                                                                        & 213$\pm$20     &                           \\ \cline{1-1} \cline{3-4} \cline{7-7}
La$_{1.82}$Sr$_{0.18}$CuO$_4$            &                                                                                   &                     & 1.2$\pm$0.4           & \multicolumn{2}{l|}{}                                                                                                                        & 203$\pm$15     &                           \\ \cline{1-1} \cline{3-4} \cline{7-7}
La$_{1.81}$Sr$_{0.19}$CuO$_4$            &                                                                                   &                     & 1.1$\pm$0.4           & \multicolumn{2}{l|}{}                                                                                                                        & 198$\pm$15     &                           \\ \cline{1-1} \cline{3-4} \cline{7-7}
La$_{1.8}$Sr$_{0.2}$CuO$_4$              &                                                                                   &                     & 1.1$\pm$0.4           & \multicolumn{2}{l|}{}                                                                                                                        & 197$\pm$15     &                           \\ \cline{1-1} \cline{3-4} \cline{7-7}
La$_{1.79}$Sr$_{0.21}$CuO$_4$            &                                                                                   &                     & 1.05$\pm$0.4          & \multicolumn{2}{l|}{}                                                                                                                        & 198$\pm$15     &                           \\ \cline{1-1} \cline{3-4} \cline{7-7}
La$_{1.78}$Sr$_{0.22}$CuO$_4$            &                                                                                   &                     & 0.9$\pm$0.4           & \multicolumn{2}{l|}{}                                                                                                                        & 199$\pm$15     &                           \\ \cline{1-1} \cline{3-4} \cline{7-7}
La$_{1.77}$Sr$_{0.23}$CuO$_4$            &                                                                                   &                     & 1.08$\pm$0.4          & \multicolumn{2}{l|}{}                                                                                                                        & 199$\pm$15     &                           \\ \cline{1-1} \cline{3-4} \cline{7-7}
La$_{1.76}$Sr$_{0.24}$CuO$_4$            &                                                                                   &                     & 1.0$\pm$0.4           & \multicolumn{2}{l|}{}                                                                                                                        & 199$\pm$15     &                           \\ \cline{1-7}
La$_{1.93}$Sr$_{0.07}$CuO$_4$            &                                                                                   & 12.3                & 4.79$\pm$0.018       & 200-400                                                                                        & \multirow{19}{*}{\cite{Ando1}}                        & 497$\pm$37     &                           \\ \cline{1-1} \cline{3-5} \cline{7-7}
La$_{1.92}$Sr$_{0.08}$CuO$_4$            &                                                                                   & 22                  & 3.96$\pm$0.016        & 200-400                                                                                        &                                             & 377$\pm$30     &                           \\ \cline{1-1} \cline{3-5} \cline{7-7}
La$_{1.91}$Sr$_{0.09}$CuO$_4$            &                                                                                   & 24.5                & 3.21$\pm$0.02         & 200-400                                                                                        &                                             & 314$\pm$30     &                           \\ \cline{1-1} \cline{3-5} \cline{7-7}
La$_{1.9}$Sr$_{0.1}$CuO$_4$              &                                                                                   & 27.5                & 2.75$\pm$0.003        & 250-400                                                                                        &                                             & 286$\pm$30     &                           \\ \cline{1-1} \cline{3-5} \cline{7-7}
La$_{1.89}$Sr$_{0.11}$CuO$_4$            &                                                                                   & 29.6                & 2.34$\pm$0.002        & 250-400                                                                                        &                                             & 282$\pm$30     &                           \\ \cline{1-1} \cline{3-5} \cline{7-7}
La$_{1.88}$Sr$_{0.12}$CuO$_4$            &                                                                                   & 30.4                & 2.05$\pm$0.002        & 250-400                                                                                        &                                             & 280$\pm$20     &                           \\ \cline{1-1} \cline{3-5} \cline{7-7}
La$_{1.87}$Sr$_{0.13}$CuO$_4$            &                                                                                   & 34.6                & 1.72$\pm$0.001        & 200-400                                                                                        &                                             & 277$\pm$20     &                           \\ \cline{1-1} \cline{3-5} \cline{7-7}
La$_{1.86}$Sr$_{0.14}$CuO$_4$            & \multirow{9}{*}{La$_{2-x}$Sr$_x$CuO$_4$}                                                     & 36.3                & 1.54$\pm$0.006        & 200-400                                                                                        &                                             & 268$\pm$20     &                           \\ \cline{1-1} \cline{3-5} \cline{7-7}
La$_{1.85}$Sr$_{0.15}$CuO$_4$            &                                                                                   & 39.3                & 1.44$\pm$0.001        & 180-400                                                                                        &                                             & 249$\pm$20     &                           \\ \cline{1-1} \cline{3-5} \cline{7-7}
La$_{1.84}$Sr$_{0.16}$CuO$_4$            &                                                                                   & 36.6                & 1.26$\pm$0.004        & 130-400                                                                                        &                                             & 229$\pm$20     &                           \\ \cline{1-1} \cline{3-5} \cline{7-7}
La$_{1.83}$Sr$_{0.17}$CuO$_4$            &                                                                                   & 35.7                & 1.18$\pm$0.004        & 200-400                                                                                        &                                             & 213$\pm$20     &                           \\ \cline{1-1} \cline{3-5} \cline{7-7}
La$_{1.82}$Sr$_{0.18}$CuO$_4$            &                                                                                   & 36                  & 1.05$\pm$0.003        & 200-400                                                                                        &                                             & 203$\pm$15     &                           \\ \cline{1-1} \cline{3-5} \cline{7-7}
La$_{1.81}$Sr$_{0.19}$CuO$_4$            &                                                                                   & 33                  & 0.99$\pm$0.006        & 50-400                                                                                         &                                             & 198$\pm$15     &                           \\ \cline{1-1} \cline{3-5} \cline{7-7}
La$_{1.8}$Sr$_{0.2}$CuO$_4$              &                                                                                   & 30.3                & 0.96$\pm$0.001        & 150-400                                                                                        &                                             & 197$\pm$15     &                           \\ \cline{1-1} \cline{3-5} \cline{7-7}
La$_{1.79}$Sr$_{0.21}$CuO$_4$            &                                                                                   & 28.5                & 0.92$\pm$0.001        & 220-400                                                                                        &                                             & 198$\pm$15     &                           \\ \cline{1-1} \cline{3-5} \cline{7-7}
La$_{1.78}$Sr$_{0.22}$CuO$_4$            &                                                                                   & 25.5                & 0.88$\pm$0.001        & 260-400                                                                                        &                                             & 199$\pm$15     &                           \\ \cline{1-5} \cline{7-8}
Bi$_2$Sr$_{1.8}$La$_{0.2}$CuO$_{6+\delta}$         & \multirow{3}{*}{Bi$_2$Sr$_{2-x}$La$_x$CuO$_{6+\delta}$}                                                & 28.1                & 1.24$\pm$0.004        & 250-300                                                                                        &                                             & 319$\pm$25     & \multirow{3}{*}{\cite{russo}}      \\ \cline{1-1} \cline{3-5} \cline{7-7}
Bi$_2$Sr$_{1.6}$La$_{0.4}$CuO$_{6+\delta}$         &                                                                                   & 29                  & 1.62$\pm$0.004        & 250-300                                                                                        &                                             & 297$\pm$20     &                           \\ \cline{1-1} \cline{3-5} \cline{7-7}
Bi$_2$Sr$_{1.4}$La$_{0.6}$CuO$_{6+\delta}$         &                                                                                   & 12                  & 3.25$\pm$0.01         & 250-300                                                                                        &                                             & 553$\pm$40     &                           \\ \hline
UPt$_3$                        & \multirow{2}{*}{UPt$_3$}                                                             & 0.5                 & 9.2$\pm$0.2           & 5-10                                                                                          & \multirow{2}{*}{\cite{Joynt}}                        & 715        & \multirow{2}{*}{\cite{Joynt}}      \\ \cline{1-1} \cline{3-5} \cline{7-7}
$c$-axis Upt$_3$                 &                                                                                   & 0.5                 & 3.3$\pm$0.12          & 5-10                                                                                          &                                             & 422        &                           \\ \hline
CeCoIn$_5$  (0 Gpa)            & \multirow{4}{*}{CeCoIn$_5$}                                                          & 2.3                 & 1.61$\pm$0.008        & 3-20                                                                                          & \multirow{4}{*}{Our data}                   & 350$\pm$12    & \multirow{4}{*}{\cite{Howald}}      \\ \cline{1-1} \cline{3-5} \cline{7-7}
CeCoIn$_5$  (0.3 Gpa)          &                                                                                   & 2.51                & 1.36$\pm$0.007       & 3-20                                                                                          &                                             & 300$\pm$12    &                           \\ \cline{1-1} \cline{3-5} \cline{7-7}
CeCoIn$_5$  (0.55 Gpa)         &                                                                                   & 2.58                & 1.20$\pm$0.002       & 3-20                                                                                          &                                             & 280$\pm$12    &                           \\ \cline{1-1} \cline{3-5} \cline{7-7}
CeCoIn$_5$  (1 Gpa)            &                                                                                   & 2.63                & 0.97$\pm$0.004       & 3-20                                                                                          &                                             & 262$\pm$12    &                           \\ \hline
\caption{Transport parameters and London penetration depth at zero temperature.}
\end{longtable}
\end{small}

\begin{figure}[htp]
\centering
\includegraphics[trim=0cm 0cm 0cm 0cm, clip=true, width=0.5\textwidth]{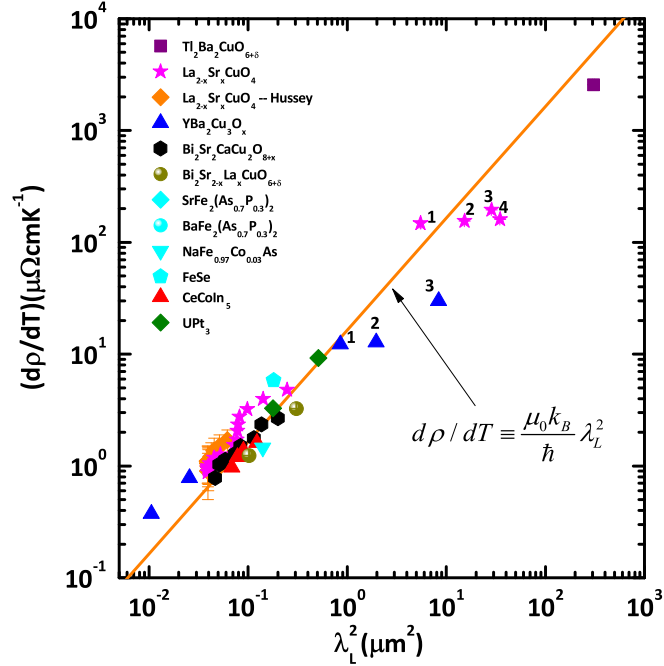}
\caption{\label{fig:TORQUEvsT} (Color online)  Log-log plot of $d\rho/dT$  vs. $\lambda^2_L$  for various strongly correlated superconductors. The orange line is the scaling relation $d\rho/dT=(\mu_0k_B/\hbar)\lambda^2_L$. See Table 1 for details, including errors.  }
\end{figure}

\section*{Conclusion}

In summary, we observed a universal scaling relation $d\rho/dT=(\mu_0k_B/\hbar)\lambda^2_L$, which connects linear-$T$-dependent resistivity to superconducting superfluid density at zero temperature in strongly correlated superconductors. Our analysis suggests that the quantum diffusion might be the origin of this scaling relation. In this case, the charge transport is viewed as a diffusion process of quasi-particles with a diffusion coefficient that approaches the quantum limit, $D\sim \hbar/m^*$.

\section*{Method}
The high quality CeCoIn$_5$ single crystal samples are grown by an indium self-flux method \cite{Petrovic}. High quality crystals  were chosen to perform the transport measurements. Four leads were attached to the single crystal, with the current applied parallel to the crystallographic $a$ axis. The resistivity was measured both in ambient pressure as well as  under  hydrostatic pressure $P$.



\section*{Acknowledgements}
We sincerely thank Prof. Mianheng Jiang, Prof. Xiaoming Xie, Dr. Wei Li, Prof. Haicang Ren, Prof. Ting-Kuo Lee, Prof. Yan Chen and Prof. Ang Li for discussions. This study was supported by the National Natural Science Foundation of China (Grant Nos. 11574338, U1530402, 11522435), the Strategic Priority Research Program (B) of the Chinese Academy of Sciences (Grant No. XDB04040300, XDB07020200) and the Youth Innovation Promotion Association of the Chinese Academy of Sciences.

\section*{Author contributions statement}
T. H. Planned the research. Y. L.  carried out the experiment. T. H. and H. X wrote the manuscript. G. M. and Y. Y. give fruitful discussions. All authors were intensively involved in the research.

\section*{Additional information}

\textbf{Competing financial interests:} The authors declare no competing financial interests.

\end{document}